\documentclass{aa}

\usepackage[varg]{txfonts}
\usepackage{natbib}
\bibpunct{(}{)}{;}{a}{}{,} 
\usepackage{multirow} 
\usepackage{array} 
\newcolumntype{x}[1]{>{\centering\arraybackslash\hspace{0pt}}p{#1}} 

\usepackage{changes} 
\begin{document}

\title{Evolution of the critical torus instability height and CME likelihood in solar active regions}

\author{Alexander W. James\inst{1} 
\and David R. Williams\inst{1} 
\and Jennifer O'Kane\inst{2} 
}

\institute{European Space Agency (ESA), European Space Astronomy Centre (ESAC), Camino Bajo del Castillo, s/n, 28692 Villanueva de la Ca{\~n}ada, Madrid, Spain
\email{alexander.james@esa.int}\label{inst1}
\and Mullard Space Science Laboratory, University College London, Holmbury St. Mary, Dorking, Surrey, RH5 6NT, UK
}

\date{Received 14 December 2021 / Accepted 21 June 2022}

\abstract
{}
{Working towards improved space weather predictions, we aim to quantify how the critical height at which the torus instability drives coronal mass ejections (CMEs) varies over time in a sample of solar active regions.}
{We model the coronal magnetic fields of 42 active regions and quantify the critical height at their central polarity inversion lines throughout their observed lifetimes. 
We then compare these heights to the changing magnetic flux at the photospheric boundary and identify CMEs in these regions.}
{In our sample, the rates of CMEs per unit time are twice as high during phases when magnetic flux is increasing than when it is decreasing, and during those phases of increasing flux, the rate of CMEs is 63\% higher when the critical height is rising than when it is falling.
Furthermore, we support and extend the results of previous studies by demonstrating that the critical height in active regions is generally proportional to the separation of their magnetic polarities through time.
When the separation of magnetic polarities in an active region increases, for example during the continuous emergence and expansion of a magnetic bipole, the critical height also tends to increase. Conversely, when the polarity separation decreases, for example due to the emergence of a new, compact bipole at the central inversion line of an existing active region or into a quiet Sun environment, the critical height tends to decrease.}
{}


\titlerunning{Evolution of critical height and CME likelihood in active regions} 
\authorrunning{A.W. James et al.}

\maketitle


\section{Introduction} \label{sec:intro}

Coronal mass ejections (CMEs) from the Sun can significantly impact human activities on Earth, in near-Earth space, and throughout the heliosphere.
However, we are presently unable to predict the eruptions of CMEs, so accurate forecasting of CME effects can only begin once an eruption has already occurred. CMEs typically take between 1 and 3 days to reach the Earth, so the production of space weather forecasts with 5-day lead times, as desired by the affected sectors \citep{krausmann2016}, would require the prediction of CMEs 2--4 days before they occur.

Additionally, improved CME predictions would directly benefit scientific missions.
Launched in February 2020 and beginning its nominal mission phase in November 2021, the European Space Agency's \textit{Solar Orbiter} mission \citep{muller2020solo} aims to answer a number of outstanding science questions, including how solar transients, such as CMEs, drive variability in the heliosphere \citep{zouganelis2020SAP}. \textit{Solar Orbiter} boasts an impressive suite of in situ and remote-sensing instruments that will collect data to help to answer this question. The \textit{Extreme Ultraviolet Imager} (EUI; \citealp{rochus2020EUI}) and the \textit{Polarimetric and Helioseismic Imager} (PHI; \citealp{solanki2020PHI}) instruments each feature multiple telescopes, with one dedicated to imaging the full solar disc, and at least one other that can image smaller, targeted fields-of-view at higher spatial resolutions (PHI has one high-resolution telescope and EUI has two for observing at extreme-ultraviolet (EUV) and Lyman-$\alpha$ wavelengths). In order to maximise the efficient collection of high-quality data and answer the proposed science questions, the targeting of \textit{Solar Orbiter}'s high-resolution remote-sensing observations must be planned meticulously, compounded also by the time constraints on planning observations for a mission in deep space. By identifying regions on the Sun that are more likely to produce eruptive events, observation planners such as those for the \textit{Solar Orbiter} mission, the \textit{EUV Imaging Spectrometer} (\citealp{culhane2007eis}) and \textit{Solar Optical Telescope} (\citealp{tsuneta2008sot}) instruments onboard \textit{Hinode} \citep{kosugi2007hinode}, and the upcoming \textit{European Solar Telescope} (\citealp{jurcak2019EST}) can optimise the yield of valuable data.

By developing our understanding of the mechanisms involved in CME initiation, we can work towards predicting CMEs hours or even days before they occur.
At present, the commonly reported mechanisms that drive CMEs into interplanetary space can be categorised into two groups: resistive and ideal processes \citep[Table 1 of][]{green2018eruptions}. On the one hand, some studies attribute CME acceleration to a runaway process of magnetic flare-reconnection, e.g. in a `magnetic breakout' scenario \citep{antiochos1999model}, whereas others conclude that CMEs are driven by an ideal magnetohydrodynamic instability, such as the torus instability \citep{kliem2006torus}. 
The torus instability occurs when the outward expansion of a toroidal structure can not be balanced by an oppositely-directed external force. In the solar atmosphere, twisted, semi-toroidal structures known as magnetic flux ropes carry electric currents and tend to expand via the hoop force, with overlying magnetic loops contributing a constraining external tension force.

A parameter known as the magnetic decay index, $n$, quantifies how the strength of the poloidal component of an external magnetic field, $B_{\mathrm{ext,p}}$, (the component that contributes to the tension force) changes with the radius of a torus, $R$:
\begin{equation}
    n = - \frac{d \ln{B_{\mathrm{ext,p}}}}{d \ln{R}}.
    \label{eqn:decay_index}
\end{equation}
\citet{bateman1978instabilities} showed that a ring of current would be unstable to expansion if embedded within an external magnetic field where the decay index is above a critical value of $n_{\mathrm{c}}=1.5$, and \citet{kliem2006torus} later demonstrated the applicability of this scenario to the driving of CMEs. 
If $n \geq n_{\mathrm{c}}$ at the axis of a flux rope, the structure will undergo runaway expansion, erupting as a CME. 
Theoretical studies have found values of $n_{\mathrm{c}}=1.1-1.3$ for straight flux ropes \citep{demoulin2010criteria} and $n_{\mathrm{c}}=1.5-1.9$ in circular flux ropes \citep{torok2007simulations,fan2007onset}. Recent observational results identify average thresholds of $n_{\mathrm{c}}=1.2\pm0.2$ and $n_{\mathrm{c}}=1.6\pm0.1$ in eruptions of quiescent filaments and hot channel flux ropes, respectively \citep{cheng2020initiation}.

It has been shown that the height at which the critical decay index occurs (hereafter referred to simply as the critical height, $h_{\mathrm{c}} = h(n=n_{\mathrm{c}})$) in active regions at the times of two-ribbon flares is approximately half of the distance between the centroids of the region's leading and trailing magnetic polarities \citep{wang2017critical,baumgartner2018eruptive}. 
In this paper, we extend this result by investigating how the critical height in active regions varies over time, not only at the onset times of eruptions, and by interpreting the results in the context of the evolving active region magnetic flux, which we break down into phases.

The magnetic flux in active regions is strongly linked to eruptive activity.
The formation of pre-eruptive structures can be triggered during periods of magnetic flux emergence \citep{james2017on-disc,james2018model} as well as flux cancellation \citep{green2011cancellation,yardley2018cancellation}, and it is widely known that more magnetically complex active regions (e.g. with $\beta\gamma$ and $\delta$ Mount Wilson classifications) produce more CMEs than magnetically simple ($\alpha$ and $\beta$ class) regions \citep{chen2011statistical}. 
By examining the changing quantity and complexity of magnetic flux in active regions, tracking the calculated critical height, and identifying the onset times of CMEs from those regions, we aim to understand how these factors in the solar atmosphere combine to create the most favourable conditions for a CME.

In Section \ref{sec:data_events}, we outline the data used in this work and the regions we selected for study. Section \ref{sec:methods} contains descriptions of the methods used for quantifying various parameters, categorising the observed lives of active regions into phases, and identifying CMEs, and Section \ref{sec:results} contains our results. We discuss our findings and their implications in Section \ref{sec:discussion} and present conclusions in Section \ref{sec:conclusions}.

\section{Data and event selection} \label{sec:data_events}

In this work, we use observations of the photospheric magnetic field from the \textit{Helioseismic and Magnetic Imager} (HMI; \citealp{scherrer2012hmi}) onboard the \textit{Solar Dynamics Observatory} (SDO; \citealp{pesnell2012SDO}). HMI Active Region Patches (HARPs; \citealp{hoeksema2014helioseismic}) provide a convenient method of providing fixed boundaries around one active region or more whilst containing the full extent of the region(s) at all times. We use radial magnetic field component magnetograms from the hmi.sharp\_cea\_720s dataset. Magnetograms in this SHARP dataset (Spaceweather HMI Active Region Patch; \citealp{bobra2014sharp}) are provided in a cylindrical equal-area projection, meaning each pixel corresponds to an equal area of the solar surface ($0.03^{\circ} \times 0.03^{\circ}$) and the magnetograms are projected in a way such that the observer's perspective is as if directly above the HARP at any longitude. We spatially rebin the magnetograms by a factor of 6, such that each pixel represents a width of $\approx 2.18\ \mathrm{Mm}$ at a distance of 1 AU, and we use magnetograms that are taken 1 hour apart.

We select 42 HARPs, spanning much of solar cycle 24, and examine one NOAA active region in each HARP. The active regions display a wide range of magnetic complexities, including periods with relatively simple $\alpha$ and $\beta$ classes, and more complex $\gamma$ and $\delta$ Mount Wilson classifications. To bolster the number of bipolar active regions in our sample, we include some regions that were previously studied by \citet{yardley2018cancellation}, and for more regions with complex magnetic structures, we incorporate some events studied by \citet{toriumi2017magnetic}. 
HMI's magnetic field observations are noisier at the solar limb than at disc centre \citep{liu2012comparison}, so we choose to only examine the active regions at times when they are within $\pm60^{\circ}$ of central meridian as viewed from Earth. In total, we examine and extrapolate 6951 magnetograms to cover the observed lives of the 42 active regions at a 1-hour cadence.
The identifying HARP numbers of the regions studied in this work, the numbers of the NOAA active regions contained in each HARP, and the start and end times of our observations can be found in Table \ref{tbl:table_events}.

\begin{table*}
    \caption{Table of events.}
 	\centering
		\begin{tabular}{lllllll}
        \hline\hline
        HARP & NOAA & Start Time (UT) & End Time (UT) & CME Onset (UT) & $h_{\mathrm{c}}$ (Mm) & CME Phase \\
        \hline
 104 & 11092 & 2010-07-31 00:00 & 2010-08-07 13:00 & 1st@07:45               & 67 $\pm 15$      & DD \\
 114 & 11095 & 2010-08-05 09:00 & 2010-08-13 00:00 &                         &         &        \\
 131 & 11098 & 2010-08-11 00:00 & 2010-08-17 23:00 &                         &         &        \\
 135 & 11100 & 2010-08-18 06:00 & 2010-08-24 20:00 &                         &         &        \\
 146 & 11102 & 2010-08-25 04:00 & 2010-09-02 00:00 &                         &         &        \\
 241 & 11120 & 2010-11-02 00:00 & 2010-11-09 00:00 &                         &         &        \\
 377 & 11158 & 2011-02-10 23:00 & 2011-02-18 00:00 & 13th@17:34, 14th@02:39, & $33 \pm 15$, $38 \pm 12$  & II, ID,\\
     &       &                  &                  & 14th@06:52, 14th@12:44, & $39 \pm 10$, $40 \pm 7$  & ID, ID,\\
     &       &                  &                  & 14th@17:24, 14th@19:23, & $41 \pm 5$, $41 \pm 5$  & ID, ID,\\
	 &       &                  &                  & 15th@01:48              & $42 \pm 3$      & ID     \\
 429 & 11173 & 2011-03-16 03:00 & 2011-03-24 02:00 & 19th@15:13              & $30 \pm 2$      & DD     \\
 451 & 11183 & 2011-03-29 14:00 & 2011-04-05 19:00 &                         &         &        \\
 538 & 11200 & 2011-04-26 08:00 & 2011-05-04 08:00 & 29th@12:46              & 28 $\pm 4$      & ID     \\
 661 & 11234 & 2011-06-10 16:00 & 2011-06-18 05:00 & 14th@04:20, 15th@03:43, & 37 $\pm 4$, 37 $\pm 5$, & ID, ID,\\
	 &       &                  &                  & 17th@00:15              & 39 $\pm 4$      & II     \\
 750 & 11261 & 2011-07-29 00:00 & 2011-08-05 06:00 & 2nd@05:40, 3rd@13:18,   & $23 \pm 7$, $33 \pm 20$, & DD, II,\\
     &       &                  &                  & 4th@03:44               & $44 \pm 28$      & II     \\
 833 & 11283 & 2011-09-02 18:00 & 2011-09-09 09:00 & 6th@01:40, 6th@22:13,   & $47 \pm 17$, $54 \pm 20$, & DD, DD,\\
     &       &                  &                  & 7th@22:34, 9th@05:59    & $43 \pm 24$, $53 \pm 26$  & ID, ID \\
1750 & 11504 & 2012-06-11 14:00 & 2012-06-18 17:00 & 13th@12:40, 14th@13:30  & 46 $\pm 2$, 49 $\pm 2$  & II, II \\
1807 & 11515 & 2012-06-29 22:00 & 2012-07-06 06:00 & 30th@06:07, 1st@14:57,  & $39 \pm 18$, $43 \pm 26$, & DI, DD,\\
     &       &                  &                  & 4th@04:30, 4th@08:04,   & $50 \pm 27$, $50 \pm 27$, & ID, ID,\\
     &       &                  &                  & 4th@12:17, 4th@19:42,   & $50 \pm 26$, $50 \pm 24$, & ID, II,\\
     &       &                  &                  & 5th@13:02, 5th@21:39    & $48 \pm 24$, $47 \pm 24$  & II, II \\
1930 & 11542 & 2012-08-09 03:00 & 2012-08-16 14:00 & 9th@11:33, 12th@23:35,  & 42 $\pm 4$, 46 $\pm 3$, & DI, DI,\\
	&       &                  &                  & 14th@00:21              & 46 $\pm 3$   & II     \\
1997 & 11561 & 2012-08-30 01:00 & 2012-09-05 05:00 &                         &         &        \\
2137 & 11598 & 2012-10-23 21:00 & 2012-11-01 00:00 & 26th@15:48              & 29 $\pm 6$      & DD     \\
2504 & 11680 & 2013-02-24 17:00 & 2013-03-04 22:00 &                         &         &        \\
3049 & 11813 & 2013-08-06 16:00 & 2013-08-11 13:00 &                         &         &        \\
3248 & 11856 & 2013-10-04 12:00 & 2013-10-12 03:00 &                         &         &        \\
3326 & 11886 & 2013-10-28 05:00 & 2013-11-02 10:00 &                         &         &        \\
3894 & 12017 & 2014-03-23 19:00 & 2014-03-31 04:00 & 28th@19:09, 28th@23:44, & 47 $\pm 9$, 47 $\pm 10$, & II, II,\\
     &       &                  &                  & 29th@17:45, 30th@11:45  & $48 \pm 15$, $48 \pm 18$  & II, II \\
3912 & 12021 & 2014-03-29 01:00 & 2014-04-04 19:00 &                         &         &        \\
3999 & 11813 & 2014-04-12 17:00 & 2014-04-19 16:00 & 17th@21:51, 18th@12:31  & 36 $\pm 5$, 40 $\pm 5$  & II, II \\
4228 & 12090 & 2014-06-13 19:00 & 2014-06-21 00:00 &                         &         &        \\
4440 & 12135 & 2014-08-07 23:00 & 2014-08-15 10:00 &                         &         &        \\
4901 & 12229 & 2014-12-05 05:00 & 2014-12-11 00:00 &                         &         &        \\
4941 & 12241 & 2014-12-16 02:00 & 2014-12-23 05:00 & 17th@01:27, 18th@21:36, & 49 $\pm 12$, 42 $\pm 9$, & II, DI,\\
	&        &                  &                  & 21st@10:59              & 48 $\pm 11$  & II     \\
5211 & 12283 & 2015-02-12 09:00 & 2015-02-15 20:00 &                         &         &        \\
5298 & 12297 & 2015-03-10 00:00 & 2015-03-16 21:00 & 9th@23:28, 11th@16:13,  & $39 \pm 15$, $44 \pm 18$, & DD, DD,\\
    &       &                  &                  & 15th@00:36              & $58 \pm 27$   & II     \\
5721 & 12374 & 2015-06-28 07:00 & 2015-06-30 06:00 &                         &         &        \\
5880 & 12401 & 2015-08-13 23:00 & 2015-08-21 15:00 & 15th@11:45              & 29 $\pm 1$      & DI     \\
5963 & 12419 & 2015-09-15 18:00 & 2015-09-23 16:00 &                         &         &        \\
5991 & 12427 & 2015-09-27 23:00 & 2015-10-05 14:00 &                         &         &        \\
6002 & 12429 & 2015-10-07 10:00 & 2015-10-15 03:00 &                         &         &        \\
6099 & 12453 & 2015-11-12 12:00 & 2015-11-18 05:00 &                         &         &        \\
6107 & 12455 & 2015-11-14 11:00 & 2015-11-21 02:00 &                         &         &        \\
6454 & 12526 & 2016-03-26 17:00 & 2016-04-03 08:00 &                         &         &        \\
6500 & 12532 & 2016-04-18 05:00 & 2016-04-25 16:00 &                         &         &        \\
6794 & 12599 & 2016-10-06 03:00 & 2016-10-13 06:00 &                         &         &        \\
6982 & 12648 & 2017-04-03 22:00 & 2017-04-11 16:00 &                         &         &        \\
		\hline
		\end{tabular}
 	\tablefoot{HARP and NOAA active region numbers for each event and the start and end times they are studied between. Onset times of observed CMEs are given, as well as the critical height ($h_{\mathrm{c}}$) at CME onset calculated from whichever extrapolation is based on the boundary magnetogram from the closest time. Evolutionary phases of active regions at CME onset are indicated by two letters, with the first denoting whether unsigned magnetic flux is increasing (I), decreasing (D), or approximately flat (F), and the second letter representing the same for the critical height.}
	\label{tbl:table_events}
\end{table*}

The coronal magnetic field external to a current-carrying flux rope is often approximated with a potential field \citep[e.g.][]{torok2007simulations,zuccarello2015critical,wang2017critical}. 
To produce models of the potential magnetic field, we use an implementation of the boundary value solution via Fourier transforms outlined by \citet{alissandrakis1981field} to extrapolate radial field component SHARP magnetograms into a volume with a force-free parameter equal to zero throughout the volume.
Pixels in the extrapolation volume are uniform in all spatial dimensions, sharing the scaling of the rebinned SHARP magnetograms. Thus, setting the upper boundary of the extrapolation volume to a height of 151 pixels above the lower boundary magnetogram, the resulting extrapolated field volume reaches a maximum altitude representative of $329\ \mathrm{Mm}$ or $0.47\ \mathrm{R}_{\sun}$ above the photosphere. 

The component of the external magnetic field relevant to the torus instability is the poloidal component ($B_{\mathrm{ext,p}}$ in Equation \ref{eqn:decay_index}), i.e. that perpendicular to the axis of a flux rope. Since we do not know where the axes of flux ropes may lie in the active region volumes studied in this work, we don't know which direction is the poloidal one. Instead, we choose the horizontal component ($B_{\mathrm{ext,p}}\approx{}B_{\mathrm{h}}=\sqrt{B_{x}^{2}+B_{y}^{2}}$) as a general approximation. This assumption has been made before in similar studies \citep[e.g.][]{liu2008instabilites,zuccarello2015critical,wang2017critical}, and also helps to account for any rotation of the flux rope axis during the onset of a CME that would cause the poloidal direction to change over time within the horizontal plane. By taking the vertical direction in our extrapolated field volumes as the direction of interest relevant to the torus instability ($R$ in Equation \ref{eqn:decay_index}), we can compute the decay index throughout the extrapolated potential field volumes.

To identify CMEs in each active region, we examine observations from the \textit{Atmospheric Imaging Assembly} (AIA; \citealp{lemen2012atmospheric}) onboard SDO using the JHelioviewer software \citep{muller2017jhelioviewer}. 
The numerous channels of AIA enable us to examine plasma at different temperature ranges, and in this work we analyse observations from the 131 \AA{} (response peaks at $10^{5.6}\ \mathrm{K}$ and $10^{7.0}\ \mathrm{K}$), 171 \AA{} ($10^{5.8}\ \mathrm{K}$), 193 \AA{} ($10^{6.2}\ \mathrm{K}$ and $10^{7.3}\ \mathrm{K}$), 211 \AA{} ($10^{6.3}\ \mathrm{K}$), 304 \AA{} ($10^{4.7}\ \mathrm{K}$), and 1600 \AA{} ($10^{5.0}\ \mathrm{K}$) channels. 
AIA images in JHelioviewer are available at the full spatial resolution of $1.5''$ and a reduced cadence of 36 seconds (in comparison to the native 12 second cadence of the instrument). 

We also check for eruption signatures using the 195 \AA{} channel of the \textit{Extreme Ultraviolet Imager} (EUVI) instrument which is part of the \textit{Sun Earth Connection Coronal and Heliospheric Investigation} (SECCHI; \citealp{howard2008secchi}) package onboard the \textit{Solar Terrestrial Relations Observatory} (STEREO; \citealp{kaiser2008stereo}) spacecraft. The STEREO spacecraft provide complementary observation angles to those from SDO, at times enabling us to view active regions simultaneously from above and from the side. The spatial resolution of EUVI is $3.2''$, and 195 \AA{} EUVI images in JHelioviewer are available at a cadence of 5 minutes.

To identify whether the eruption signatures seen in AIA and EUVI observations are associated with successful eruptions of CMEs, we inspect white light images from the C2 telescope of the \textit{Large Angle and Spectrometric Coronagraph} (LASCO; \citealp{brueckner1995large}) onboard the \textit{Solar and Heliospheric Observatory} (SOHO; \citealp{domingo1995soho}) and the COR2 coronagraphs from STEREO SECCHI. LASCO C2 observes between $1.5$ and $6\ \mathrm{R}_{\sun}$ from disc centre at a cadence of 12 minutes, and the STEREO COR2 telescopes image from $2.5-15\ \mathrm{R}_{\sun}$ at a cadence of 15 minutes.

Full details of the method used to identify CMEs in the studied active regions are given in Section \ref{sec:method_cmes}. The onset times of the observed CMEs are presented in Table \ref{tbl:table_events}, corresponding to either the start time of the eruptive flare or the fast rise of erupting structures. We also record the critical height at CME onset calculated from the extrapolation of whichever magnetogram was imaged at the closest time, and indicate the evolutionary phase the source active region was in at the time of each CME, based on the changing unsigned magnetic flux and critical height (see Section \ref{sec:method_phases} for more information on how these phases are determined).

\section{Methods} \label{sec:methods}

\subsection{Quantifying magnetic flux, polarity separation, and critical height} \label{sec:method_quantify}

We calculate the magnetic flux crossing the photosphere in each studied active region at every observed time step. If the SHARP magnetograms for a given event feature more than one active region, we only count magnetic flux within a cropped area that contains the active region we are interested in. 

\begin{figure}
 \centering
 \resizebox{\hsize}{!}{\includegraphics{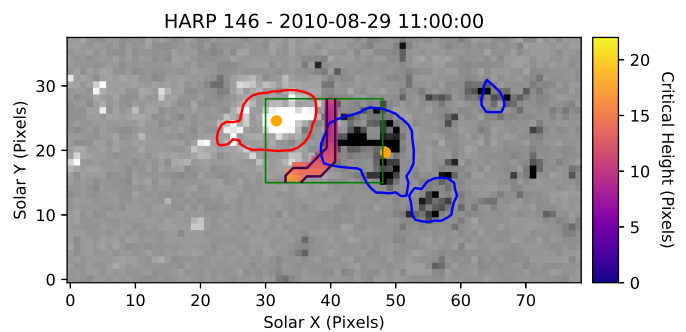}}
 \caption{Red and blue contours outline the pixels used for totalling the positive and negative radial magnetic flux, respectively. Two orange circles indicate the flux-weighted centres of the contoured positive and negative magnetic flux. The green rectangle shows the user-defined region of interest in which polarity inversion lines are detected, and inversion line pixels are coloured according to the local critical height. Pixels represent approximately $2.18\ \mathrm{Mm}$ in each spatial dimension.}
 \label{fig:pil_contours}
\end{figure}

To remove the significant contribution of magnetic flux that comes from noisy measurements in weak-field pixels \citep{hoeksema2014helioseismic}, we produce a mask by first applying a $7\times7$ pixel spatial smoothing to the magnetogram, and then excluding pixels where the smoothed field strength is below $100\ \textrm{G}$. We then apply this mask to the original magnetogram and sum the unmasked positive and negative magnetic flux. The thresholding and masking is illustrated in Figure \ref{fig:pil_contours} with contours that outline the pixels to be included in flux calculations. The unsigned magnetic flux is computed by adding the magnitudes of the positive and negative fluxes together and dividing by two.

From the masks of magnetic flux described above, we also find the flux-weighted centroids of positive and negative polarity in each SHARP magnetogram. We use the distance between these centroids to quantify the mean separation between positive and negative polarities in each region through time. Whilst this method provides an effective representation of polarity separation in relatively bipolar active regions, it may be less representative in multipolar regions.

Although we compute the decay index throughout the full extrapolated potential field volumes, the critical height for CMEs is only relevant at the locations where magnetic flux ropes may form and erupt. Flux ropes have their footpoints in opposite magnetic polarities, and their axes tend to lie along polarity inversion lines (PILs), evidenced by filament channels in some lower-lying flux ropes \citep[e.g.][]{wang2019signatures} and even in nonlinear force-free field extrapolations of higher altitude flux ropes \citep[e.g.][]{james2018model}. Therefore, we focus our analysis of the critical height along PILs. Practically, we define a rectangular sub-region of interest within each magnetogram, and then detect PILs in that area automatically by finding neighbouring pixels where the polarity of the smoothed radial magnetic field component changes sign (illustrated in Figure \ref{fig:pil_contours}). The sub-regions are chosen so that they completely encompass the PIL that exists between the strong magnetic polarities in each active region. This is intended to avoid examining parts of the PIL that are ill-defined in the periphery of the active region where the magnetic field is weaker and noisier. We then find the height above each PIL pixel at which the decay index changes from sub-critical to critical and take the mean of these critical heights along the PIL to give a single, average critical height that is representative for each active region at a given time. We estimate the uncertainty in the critical heights using the standard deviation of the mean. At times, there may be multiple instances of the critical height above a single pixel (i.e. the decay index profile in height is saddle shaped; \citealp{guo2010confined,luo2022saddles}). We remove the bottom three layers of each extrapolation volume in order to neglect spurious critical heights that can manifest there as a result of noise in the boundary photospheric magnetograms, and then we select the lowest remaining critical height above each pixel. In some cases, saddle-profiles have been reported to lead to confined or two-step eruptions when erupting structures encounter a torus-stable region between two critical heights \citep[e.g.][]{gosain2016interrupted,liu2018failed}, but successful eruptions can also occur \citep[e.g.][]{wang2017critical} when the erupting structure continues to accelerate through the stable region due to the feedback process of magnetic reconnection \citep{inoue2018modeling}.

\subsection{Categorising phases of active region evolution} \label{sec:method_phases}

Having followed the procedures outlined in Section \ref{sec:method_quantify}, we have values of the total unsigned magnetic flux, average critical height, and mean magnetic polarity separation at a 1 hour cadence throughout the lives of each HARP region. To remove the systematic effects of HMI's 12- and 24-hour periodic variations in observed magnetic flux \citep{liu2012comparison,hoeksema2014helioseismic}, we smooth the curves of magnetic flux, critical height, and polarity separation in time using a moving average window of 24 data points. Since this method can not be applied equally at the boundaries of the curves (i.e. within the first and last 12 data points of each sequence), the ($N<12$)\textsuperscript{th} data point is replaced with the average of itself, the preceding N data points, and the subsequent 11 points. Similarly, the ($N>L-12$)\textsuperscript{th} data point (where L is the total number of data points in the series) is replaced by the average of itself, the preceding 12 data points, and the subsequent $L-N$ points. Plainly, the smoothing of quantities is asymmetric near the beginning and end of each time series, using fewer data points.

We divide the lives of each active region into several `phases' based on intervals of time during which the (smoothed) unsigned magnetic flux and critical height are evolving uniformly. Specifically, we categorise each phase of an active region's observed life in to one of nine categories, based on whether the magnetic flux is increasing, decreasing, or flat (unchanging), and whether the critical height is increasing, decreasing, or flat over a given time period. 
Due to the ever-evolving nature of the solar atmosphere, we rarely see the unsigned flux and critical height remain truly `flat' for more than a couple of data points (within the precision of 1/100th of the total unsigned flux and 1/10th of a Mm), however we also assign the `flat' label if either quantity tends to oscillates around a central value for several data points, with no significant net increase or decrease. 
When the increasing/decreasing/flat trend of one or both of the unsigned flux and critical height changes, we mark the boundary of a new phase.
In cases where the trends of the flux and critical height both change within just a few hours of each other, we designate only one new phase rather than defining two new phases shortly after each other, with the boundary time marked as the middle time between the turning points. After all, there is some room for uncertainty in determining the precise times at which new phases begin because we smooth the observed quantities in time, as described in the previous paragraph, and because there can be a lag of several hours between the emergence of a new bipole and the reconfiguration of coronal field at high altitudes that determines the critical height.

By defining phases, we can collect statistics on how often increasing/decreasing/flat magnetic flux evolution in an active region is accompanied by an increasing/decreasing/flat critical height. To ensure a comprehensive and balanced study, our sample of active regions was selected to capture a mixture of magnetic complexities. To see the effect of magnetic complexity on our results, we further subdivide these phases according to the Mount Wilson class of each active region. This allows us to distinguish between, for example, periods of increasing magnetic flux and increasing critical height in `simple' ($\alpha$ or $\beta$ class) and `complex' ($\gamma$ or $\delta$ class) active regions. Whilst the magnetic complexity of an active region can physically vary on timescales of minutes (e.g. due to magnetic flux emergence), the Mount Wilson class of an active region is only reassessed daily. This means the categorisation of phases by their complexity is subject to some uncertainty in time. On the other hand, the Mount Wilson scheme is an established and practical measure of complexity, and therefore its use in this work lends insight into how our methodology could be employed in space weather forecasting using existing datasets.

\subsection{Identifying CMEs} \label{sec:method_cmes}

In order to understand how the changing magnetic flux and critical height affect CME productivity, we must identify CMEs that erupt from our sample of active regions during the period of study (i.e. when the regions are between $\pm60^{\circ}$ of disc centre as viewed from SDO).
First, we identify flaring activity in EUV images, using data from the 131 \AA{}, 171 \AA{}, and 193 \AA{} channels of SDO/AIA and the 195 \AA{} channel of STEREO/EUVI. 
Then, we search for associated ejecta in the SOHO/LASCO C2 coronagraph and/or the STEREO COR2 coronagraphs. We are able to determine whether ejecta are associated with active region flares by comparing the quadrant of the corongraph images they appear in and their speeds as noted in the LASCO CME catalog\footnote{\url{https://cdaw.gsfc.nasa.gov/CME_list/}} to the location and timing of the flare on the solar disc.
The onset times of CMEs presented in Table \ref{tbl:table_events} are taken as the start time of the eruptive flare associated with each CME.

To focus on CMEs where the torus instability may indeed be a relevant driving mechanism, we select only eruptions that exhibit indicators that a magnetic flux rope may be present. These include one or more of: EUV dimmings associated with decreases in plasma density that can occur at the footpoints of erupting flux ropes \citep{thompson2000dimmings}, the eruption of cool filament material supported by concave-up sections of twisted magnetic field in a flux rope \citep{gibson2006prominence}, J-shaped flare ribbons that hook around the feet of flux ropes \citep{janvier2014electric}, teardrop shaped cavities \citep{gibson2006prominence}, and signs of the classic three-part structure of a flux rope CME in white light coronagraph images \citep{illing1985transient}.
We use base-difference processed 211 \AA{} AIA images to identify coronal dimmings, 304 \AA{} AIA images to look for filaments, 1600 \AA{} AIA images to examine flare ribbons, 195 \AA{} EUVI images to find coronal cavities, and white light observations from SOHO/LASCO C2 and STEREO COR2 to study the structures of CMEs.
Jet-like eruptions, exhibiting thin angular widths in EUV corona and white light coronagraph observations, are not included in our final list of CMEs because these eruptions are thought to be initiated by magnetic reconnection rather than an ideal instability \citep{shibata1992jets,wyper2018jets}.

\section{Results} \label{sec:results}

To understand how CME productivity changes throughout the life of an active region, we compare the critical height to the total magnetic flux over time and record the times of CMEs. We identify a number of trends that occur often across the sample of regions studied, linking the changing magnetic flux to changes in the critical height. Furthermore, we examine the correlation between the critical height and mean separation of magnetic polarities in active regions.

\subsection{Phases of magnetic flux and critical height evolution}

\begin{figure}
 \centering
 \resizebox{\hsize}{!}{\includegraphics{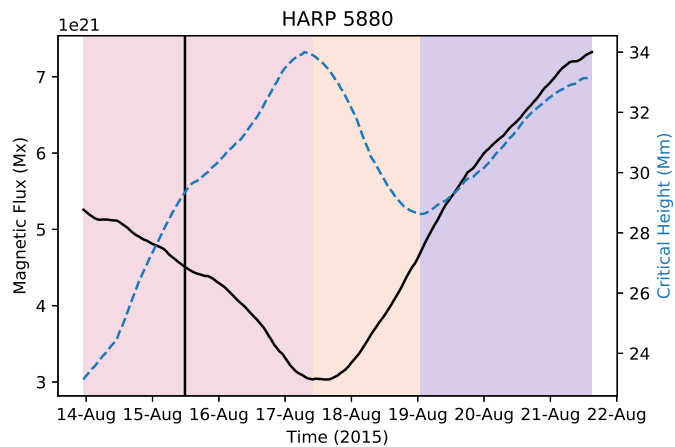}}
 \caption{Unsigned active region magnetic flux (solid black curve) and critical height (dashed blue curve) in HARP 5880 against time. Initially, the magnetic flux decreases and the critical height increases (phase shaded in pink), then magnetic flux begins to increase and the critical height falls (shaded orange). As flux emergence continues, the critical height increases once more (shaded purple). The onset time of the CME observed in this region is given by the vertical black line.}
 \label{fig:flux_vs_hc}
\end{figure}

The observed lives of the active regions are divided into distinct phases based on whether the magnetic flux and the critical height are increasing, decreasing, or relatively flat, as detailed in Section \ref{sec:method_phases} and illustrated in Figure \ref{fig:flux_vs_hc}.

We observe 2362 hours where both the unsigned magnetic flux and critical height are increasing, 2066 hours where the unsigned magnetic flux is decreasing whilst the critical height increases, 1237 hours during which the unsigned flux increases and the critical height falls, and 1034 hours where both the unsigned flux and critical height decrease. Additionally, there are 252 hours during which either the unsigned flux or the critical height does not consistently increase or decrease (they are approximately `flat' for some time). 
These values are presented in Table \ref{tbl:flux_crit_hours}.

\begin{table}
\caption{Total number of hours across the sample of 42 active regions that showed increasing, decreasing, or flat unsigned magnetic flux and critical height.}
 	\centering
\begin{tabular}{x{0.8cm}x{0.8cm}x{0.8cm}x{0.8cm}x{0.8cm}x{0.8cm}}
\hline\hline
\multicolumn{2}{c}{No. of} & \multicolumn{4}{c}{Critical Height} \\
\multicolumn{2}{c}{Hours}    & Inc.   & Dec.  & Flat  & All      \\
\hline
                    & Inc.   &  2362  & 1237  &  141  &  3740    \\
       Mag.         & Dec.   &  2066  & 1034  &   47  &  3147    \\
       Flux         & Flat   &  29    &   35  &    0  &  64      \\
                    & All    &  4457  & 2306  &  188  & 6951     \\
\hline
\end{tabular}
\label{tbl:flux_crit_hours}
\end{table}

Across the data set of 42 active regions, we identify 56 individual phases with both increasing unsigned magnetic flux and increasing critical height, 41 phases of decreasing unsigned flux and increasing critical height, 40 phases of increasing flux and decreasing critical height, and 25 phases of decreasing unsigned flux and critical height. Accounting for the total numbers of hours in each of these phase types, we see that the type of phase with the longest average duration is those with decreasing unsigned magnetic flux and increasing critical height, with a mean of 50.4 hours per phase. Second longest are the increasing flux and increasing critical height phases, at 42.2 hours, and in third, the phases with decreasing unsigned flux and decreasing critical height, at 41.4 hours. The shortest phase type is the kind with increasing unsigned flux and decreasing critical height, lasting 30.9 hours on average.

For completeness, there were 6 phases of increasing flux and flat critical height, 2 phases of decreasing flux and flat critical height, 1 phase of flat flux with increasing critical height, 2 phases of flat flux and decreasing critical height, and no phases where both the flux and critical height were flat. We do not include the mean durations of these phases in the comparison above due to the relatively small number of phases and hours available.

\subsection{Observed CMEs}

In total, we identify 47 CMEs during the studied lives of the active regions that match the observational criteria outlined in Section \ref{sec:method_cmes}.
We observe 25 CMEs during phases where both the unsigned magnetic flux and the critical height are increasing, 9 CMEs at times where the unsigned flux is decreasing and the critical height is increasing, 8 CMEs when magnetic flux is increasing and critical height is decreasing, and 5 CMEs when both the unsigned flux and critical height are decreasing. No CMEs are observed during phases where either the unsigned magnetic flux or the critical height is `flat'.
This information is presented in Table \ref{tbl:flux_crit_cmes}.
These data show more than twice as many CMEs occurring during phases of increasing magnetic flux than in phases of decreasing flux, and more than two and a half times as many CMEs at times when the critical height is increasing than when it is decreasing.

\begin{table}
\caption{Number of CMEs erupting during phases of increasing, decreasing, or flat total magnetic flux and critical height.}
 	\centering
\begin{tabular}{x{0.8cm}x{0.8cm}x{0.8cm}x{0.8cm}x{0.8cm}x{0.8cm}}
\hline\hline
\multicolumn{2}{c}{No. of} & \multicolumn{4}{c}{Critical Height}   \\
\multicolumn{2}{c}{CMEs}    & Inc.  & Dec.  & Flat   & All   \\
\hline
                    & Inc.  & 25    &  8    &  0     & 33    \\
            Mag.    & Dec.  &  9    &  5    &  0     & 14    \\
            Flux    & Flat  &  0    &  0    &  0     &  0    \\
                    & All   & 34    & 13    &  0     & 47    \\  
\hline
\end{tabular}
\label{tbl:flux_crit_cmes}
\end{table}

\begin{figure}
\centering
\resizebox{\hsize}{!}{\includegraphics{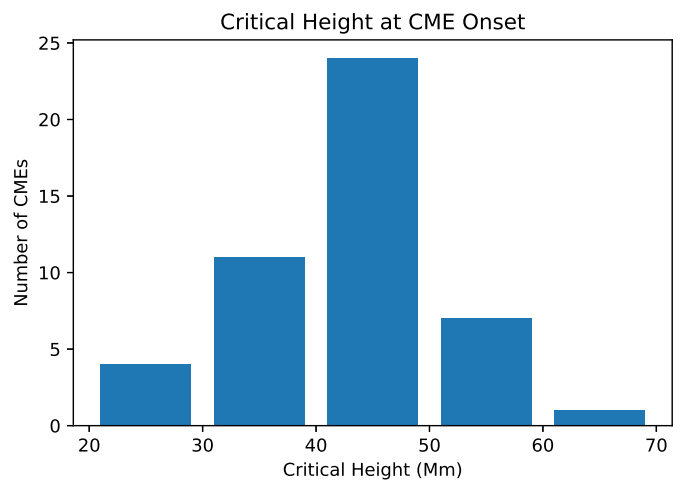}}
\caption{Distribution of critical heights recorded at the onset times of CMEs using $10\ \mathrm{Mm}$ bin widths.}
\label{fig:hc_bars}
\end{figure}

We record the critical height at the onset time of each CME in Table \ref{tbl:table_events}, and visualise the full distribution of values in Figure \ref{fig:hc_bars}. 
The critical height at CME onset is most commonly between $40\ \mathrm{Mm}$ and $50\ \mathrm{Mm}$, with a mean value $43 \pm 8\ \mathrm{Mm}$, although values as high as $67 \pm 15\ \mathrm{Mm}$ and as low as $23 \pm 7\ \mathrm{Mm}$ are found.
We find no significant dependence of the mean critical height at CME onset on the phase type of the source active region, with a mean critical height at CME onset of $44 \pm 6\ \mathrm{Mm}$ in phases of increasing flux and increasing critical height, $42 \pm 6\ \mathrm{Mm}$ when flux is decreasing and critical height is increasing, $42 \pm 7\ \mathrm{Mm}$ when flux is increasing and critical height is decreasing, and $40 \pm 15\ \mathrm{Mm}$ when both flux and critical height are decreasing.

\subsection{CMEs per phase hour}

The CME productivity of an active region can be thought of in terms of its whole lifetime, but in operational and observational scientific contexts, it is also interesting to know how frequently a region will produce CMEs in any given interval of time.
When we look at the phases in this way, we observe 1.06 CMEs per 100 hours during phases where both the unsigned magnetic flux and the critical height are increasing, 0.44 CMEs per 100 hours in phases of decreasing magnetic flux and increasing critical height, 0.65 CMEs per 100 hours when magnetic flux is increasing and critical height is decreasing, and 0.48 CMEs per 100 hours when both the magnetic flux and critical height are decreasing.
As stated in the previous section, no CMEs are observed during phases where the unsigned magnetic flux or critical height evolve flatly.
These values are presented in Table \ref{tbl:flux_crit_cmes_per_hour}.

\begin{table}
\caption{Average rate of CMEs per 100 hours during phases of increasing, decreasing, or flat unsigned magnetic flux and critical height.}
 	\centering
\begin{tabular}{x{0.8cm}x{0.8cm}x{0.8cm}x{0.8cm}x{0.8cm}x{0.8cm}}
\hline\hline
\multicolumn{2}{c}{CMEs per} & \multicolumn{4}{c}{Critical Height} \\
\multicolumn{2}{c}{100 Hours}          & Inc.   & Dec.   & Flat    & All         \\
\hline
                               & Inc.  & 1.06   & 0.65   & 0.00    & 0.88        \\
            Mag.               & Dec.  & 0.44   & 0.48   & 0.00    & 0.44        \\
            Flux               & Flat  & 0.00   & 0.00   & 0.00    & 0.00        \\
                               & All   & 0.76   & 0.56   & 0.00    & 0.68        \\
\hline
\end{tabular}
\tablefoot{These rates are obtained as ratios of the numbers of CMEs from Table \ref{tbl:flux_crit_cmes} to the numbers of (100) hours from Table \ref{tbl:flux_crit_hours}.}
\label{tbl:flux_crit_cmes_per_hour}
\end{table}

We see that twice as many CMEs erupt on average per unit time during phases of increasing unsigned magnetic flux than when flux is decreasing (0.88 vs 0.44 CMEs per 100 hours). Furthermore, there are 36\% more CMEs per unit time during phases where the critical height is increasing than when it is decreasing (0.76 vs 0.56 CMEs per 100 hours).
The phase type with the highest rate of CMEs is those with increasing unsigned magnetic flux and increasing critical height (1.06 CMEs per 100 hours), and the lowest CME rate occurs during phases of decreasing magnetic flux and increasing critical height (0.44 CMEs per 100 hours). The CME rate is almost as low in phases of decreasing flux and decreasing critical height (0.48 CMEs per 100 hours), and moderate CME rates are observed in phases where the unsigned magnetic flux increases and the critical height decreases (0.65 CMEs per 100 hours).

\subsection{The effect of magnetic complexity}

As introduced in Section \ref{sec:method_phases}, we further sub-categorise our results according to the Mount Wilson class of each active region to examine the effect of magnetic complexity on our results. If an active region has a class of $\alpha$ or $\beta$ on a given day, we classify the phase hours and CMEs from that day as occurring in a magnetically simple regime, whereas on days where an active region has a $\gamma$ or $\delta$ classification, phase hours and CMEs are categorised as magnetically complex. There are some days on which a studied active region does not have a designated Mount Wilson class, and phase hours and CMEs from these periods are counted separately.

On days where the active regions were magnetically simple (complex), we studied 1019 (991) hours where both the unsigned magnetic flux and critical height were increasing, 1230 (636) hours in which magnetic flux decreases whilst the critical height increases, 558 (433) hours where the magnetic flux increases alongside a decreasing critical height, and 637 (332) hours where both the unsigned magnetic flux and the critical height are decreasing. There are a total of 145 (53) hours where one or both of the unsigned magnetic flux and critical height are approximately flat, and a total of 917 hours where the active regions have no designated class.

During these magnetically simple (complex) periods, we observe 3 (22) CMEs in phases with increasing unsigned flux and increasing critical height, 2 (7) CMEs in phases of decreasing magnetic flux and increasing critical height, 2 (6) CMEs during periods of increasing flux and decreasing critical height, and 1 (4) CMEs when both the unsigned magnetic flux and critical height are decreasing. There are no CMEs during periods of flat magnetic flux and/or critical height evolution, regardless of magnetic complexity, and no CMEs observed from active regions at times when they do not have a defined Mount Wilson class. This information is presented in Tables \ref{tbl:flux_crit_simple} \& \ref{tbl:flux_crit_complex}.

\begin{table}
\caption{Total number of hours, number of CMEs, and rate of CMEs per 100 hours during phases of increasing, decreasing, or flat unsigned magnetic flux and critical height in magnetically simple ($\alpha$ or $\beta$) active regions.}
 	\centering
\begin{tabular}{x{0.8cm}x{0.8cm}x{0.8cm}x{0.8cm}x{0.8cm}x{0.8cm}}
\hline\hline
\multicolumn{2}{c}{No. of} & \multicolumn{4}{c}{Critical Height} \\
\multicolumn{2}{c}{Hours}    &  Inc.  & Dec.  & Flat  & All           \\
\hline
                    & Inc.   &  1019  & 558  & 87   & 1664    \\
       Mag.         & Dec.   &  1230  & 637  &  3   & 1870    \\
       Flux         & Flat   &    29  &  26  &  0   &   55    \\
                    & All    &  2278  & 1221 & 90   & 3589    \\
\end{tabular}
\begin{tabular}{x{0.8cm}x{0.8cm}x{0.8cm}x{0.8cm}x{0.8cm}x{0.8cm}}
\hline\hline
\multicolumn{2}{c}{No. of} & \multicolumn{4}{c}{Critical Height}   \\
\multicolumn{2}{c}{CMEs}    & Inc.  & Dec.  & Flat   & All   \\
\hline
                    & Inc.  &  3    &  2    &    0   &  5    \\
            Mag.    & Dec.  &  2    &  1    &    0   &  3    \\
            Flux    & Flat  &  0    &  0    &    0   &  0    \\
                    & All   &  5    &  3    &    0   &  8    \\  
\end{tabular}
\begin{tabular}{x{0.8cm}x{0.8cm}x{0.8cm}x{0.8cm}x{0.8cm}x{0.8cm}}
\hline\hline
\multicolumn{2}{c}{CMEs per} & \multicolumn{4}{c}{Critical Height} \\
\multicolumn{2}{c}{100 Hours}          & Inc.   & Dec.   & Flat   & All         \\
\hline
                               & Inc.  & 0.29   & 0.36   & 0.00   & 0.30        \\
            Mag.               & Dec.  & 0.16   & 0.16   & 0.00   & 0.16        \\
            Flux               & Flat  & 0.00   & 0.00   & 0.00   & 0.00        \\
                               & All   & 0.22   & 0.25   & 0.00   & 0.22        \\
\hline
\end{tabular}
\tablefoot{The rates of CMEs per 100 hours in the bottom part of the table are obtained as ratios of the numbers of CMEs from the middle part of the table to the numbers of (100) hours from the top section of the table.}
\label{tbl:flux_crit_simple}
\end{table}

\begin{table}
\caption{Total number of hours, number of CMEs, and rate of CMEs per 100 hours during phases of increasing, decreasing, or flat unsigned magnetic flux and critical height in magnetically complex ($\gamma$ or $\delta$) active regions.}
 	\centering
\begin{tabular}{x{0.8cm}x{0.8cm}x{0.8cm}x{0.8cm}x{0.8cm}x{0.8cm}}
\hline\hline
\multicolumn{2}{c}{No. of} & \multicolumn{4}{c}{Critical Height} \\
\multicolumn{2}{c}{Hours}    &  Inc.  & Dec.  & Flat  & All           \\
\hline
                    & Inc.   &   991  & 433  &   26  & 1450    \\
       Mag.         & Dec.   &   636  & 332  &   19  &  987    \\
       Flux         & Flat   &     0  &   8  &    0  &    8    \\
                    & All    &  1627  & 773  &   45  & 2445    \\
\end{tabular}
\begin{tabular}{x{0.8cm}x{0.8cm}x{0.8cm}x{0.8cm}x{0.8cm}x{0.8cm}}
\hline\hline
\multicolumn{2}{c}{No. of} & \multicolumn{4}{c}{Critical Height}   \\
\multicolumn{2}{c}{CMEs}    &  Inc. & Dec.  & Flat   & All   \\
\hline
                    & Inc.  &  22   &   6   &    0   &  28   \\
            Mag.    & Dec.  &   7   &   4   &    0   &  11   \\
            Flux    & Flat  &   0   &   0   &    0   &   0   \\
                    & All   &  29   &  10   &    0   &  39   \\  
\end{tabular}
\begin{tabular}{x{0.8cm}x{0.8cm}x{0.8cm}x{0.8cm}x{0.8cm}x{0.8cm}}
\hline\hline
\multicolumn{2}{c}{CMEs per} & \multicolumn{4}{c}{Critical Height} \\
\multicolumn{2}{c}{100 Hours}          & Inc.   & Dec.   & Flat   & All         \\
\hline
                               & Inc.  & 2.22   & 1.39   & 0.00   & 1.93        \\
            Mag.               & Dec.  & 1.10   & 1.20   & 0.00   & 1.11        \\
            Flux               & Flat  & 0.00   & 0.00   & 0.00   & 0.00        \\
                               & All   & 1.78   & 1.29   & 0.00   & 1.60        \\
\hline
\end{tabular}
\tablefoot{The rates of CMEs per 100 hours in the bottom part of the table are obtained as ratios of the numbers of CMEs from the middle part of the table to the numbers of (100) hours from the top section of the table.}
\label{tbl:flux_crit_complex}
\end{table}

Although we observed more hours of data in which the active regions have simple magnetic configurations, there are far more CMEs from magnetically complex active regions (8 CMEs in 3589 hours vs 39 CMEs in 2445 hours, respectively). This naturally leads to the expected result that more CMEs erupt from magnetically complex active regions per unit time than from simple active regions (1.60 vs 0.22 CMEs per 100 hours). 
In fact, the CME rates per 100 hours are systematically higher for each type of phase when active regions are complex than when they are simple.

At times when active regions are magnetically simple, we see higher rates of CMEs when unsigned magnetic flux is increasing rather than decreasing (0.30 vs 0.16 CMEs per 100 hours), and when the critical height is decreasing rather than increasing (0.25 vs 0.22 CMEs per 100 hours). The type of phase with the highest rate of CMEs is that with increasing magnetic flux and decreasing critical height (0.36 CMEs per 100 hours), and the joint lowest are when magnetic flux is decreasing and the critical height is either increasing or decreasing (0.16 CMEs per 100 hours in each).

At times when regions are magnetically complex, we see much higher rates of CMEs when unsigned magnetic flux is increasing than when it is decreasing (1.93 vs 1.11 CMEs per 100 hours), and when the critical height is increasing rather than decreasing (1.78 vs 1.29 CMEs per 100 hours). The phase type with the highest rate of CMEs is that with increasing magnetic flux and increasing critical height (2.22 CMEs per 100 hours), and the lowest is when magnetic flux is decreasing and the critical height is increasing (1.10 CMEs per 100 hours).

\subsection{Critical height vs polarity separation}

\begin{figure}
 \centering
 \resizebox{\hsize}{!}{\includegraphics{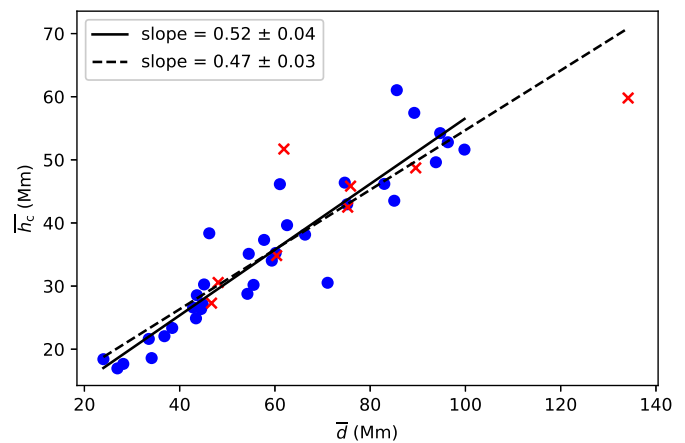}}
 \caption{Mean critical height ($\overline{h_{\mathrm{c}}}$) against mean separation between magnetic polarities ($\overline{d}$), where both quantities are averaged over the observed lives of each active region. 34 of the regions are relatively bipolar (blue circles), and as such, we expect the determined polarity separations to be accurate, whereas 8 regions are significantly multipolar (red crosses), which may cause inaccuracies in calculating the mean separation of polarities. Two linear fits are provided; one to all 42 data points (dashed line, slope = $0.47 \pm 0.03$, y-intercept = $7.42 \pm 2.25$), and one to only the 37 bipolar regions (solid line, slope = $0.52 \pm 0.04$, y-intercept = $4.55 \pm 2.27$).}
 \label{fig:hc_vs_sep}
\end{figure}

We take the time-averages of the critical height and polarity separation over the observed life of each active region and plot them against each other in Figure \ref{fig:hc_vs_sep}. To study the relationship between these quantities, we use a least-squares method to fit linearly to the data. Fitting to all 42 events, we find a slope of $0.47 \pm 0.03$, and fitting to only the 34 bipolar events where we are confident in the polarity separation method gives a slope of $0.52 \pm 0.04$. This suggests that the critical height in an active region is, on average, approximately equal to half of the separation between magnetic polarities.

\section{Discussion} \label{sec:discussion}

Of the two quantities we examine here, we find that the evolution of magnetic flux in an active region is the stronger indicator of CME likelihood per unit time. CMEs were twice as likely to occur per unit time during phases of increasing magnetic flux than during phases of decreasing magnetic flux (0.88 vs 0.44 CMEs per unit time). There also seems to be a weaker preference for CME occurrence during phases when the critical height is increasing than when it is decreasing (0.76 vs 0.56 CMEs per 100 hours). At times when the magnetic flux is decreasing, the CME rate is actually slightly higher when the critical height is decreasing than when it is increasing (0.48 vs 0.44 CMEs per 100 hours), but at times of increasing magnetic flux, there are significantly more CMEs when the critical height is rising than when it is falling (1.06 vs 0.65 CMEs per 100 hours; a difference of 63\%). This suggests that the evolution of magnetic flux could be used as a primary indicator of the predicted CME rate, and during phases of increasing flux, the critical height could serve as a secondary predictor.

In magnetically simple active regions, we see relatively few CMEs, with more CMEs occurring per unit time when unsigned magnetic flux increases rather than decreases, and when the critical height decreases rather than increases. However, due to the small sample number of CMEs, these average CME rates may be somewhat unreliable.
Magnetically complex active regions produce many CMEs, with significantly more ejections per unit time when magnetic flux is increasing rather than decreasing, and when the critical height is increasing rather than decreasing.

Across the full data sample, the type of phase that produced the most CMEs per unit time is when both the magnetic flux and the critical height are increasing, with an average of 1.06 CMEs occurring per 100 hours of this type. For the onset of the torus instability in a flux rope where the local critical height is rising, the rope must either form above the critical height, or rise sufficiently quickly in order to reach the increasing height of instability. These scenarios could indeed occur during phases of increasing magnetic flux, because flux emergence can trigger the rise of a low-lying magnetic flux rope \citep{chen2000emerging} or (in combination with with photospheric motions) trigger the formation of a flux rope at a relatively high altitude in the corona \citep{james2020trigger}. It is also possible that another triggering process, such as the helical kink instability \citep{torok2005kink} helps flux ropes to rise above the critical height, or that these CMEs are driven by mechanisms other than the torus instability.

The scenario with the second highest rate of CME occurrence per unit time in this study is the case where the magnetic flux is increasing and the critical height is decreasing, with an average of 0.65 CMEs per 100 hours. In these phases, the torus instability could drive CMEs if the critical height drops to the height of the central axis of a flux rope.

Phases where the total unsigned magnetic flux and critical height are both decreasing have the third highest CME rate, at 0.48 CMEs per 100 hours. CMEs in these types of phases could occur when the cancellation of magnetic flux at a PIL builds a low-lying flux rope, as described by \citet{van1989formation}, and the critical height falls until it meets the flux rope and initiates a CME.

Finally, phases of decreasing magnetic flux and increasing critical heights produced the fewest CMEs per unit time, at a rate of 0.44 CMEs per 100 hours. As mentioned previously, the decreasing magnetic flux could be compatible with the formation of a low-lying magnetic flux rope via flux cancellation, and such a low-lying flux rope would have to undergo a significantly rapid rise phase in order to reach the increasing critical height.

We note that six of the CMEs occurring during times of increasing unsigned flux and increasing critical height came from one particularly active 24-hour period in the magnetically complex ($\beta\gamma\delta$) NOAA active region 11158. If we were to exclude these CMEs from our analysis, the average rate of CMEs from this type of phase in complex active regions decreases from 2.22 to 1.61 CMEs per 100 hours, and from all complexities of active regions, decreases from 1.06 to 0.80 CMEs per 100 hours. In both of these cases, this type of phase continues to have the highest CME rate, and the broader trends remain, too, albeit weaker. There would still be more CMEs per unit time occurring during phases of increasing flux than decreasing flux (1.52 vs 1.11 CMEs per 100 hours for complex active regions and 0.72 vs 0.44 CMEs per 100 hours across all regions), and more CMEs per unit time when the critical height is increasing than when it is decreasing (1.41 vs 1.29 CMEs per 100 hours in complex regions and 0.63 vs 0.56 CMEs per 100 hours for all regions).

We find a mean critical height at CME onset of $43 \pm 8\ \mathrm{Mm}$.
For comparison, \citet{wang2017critical} and \citet{baumgartner2018eruptive} found mean critical heights of $36.3 \pm 17.4\ \mathrm{Mm}$ and $20.9 \pm 9.5\ \mathrm{Mm}$, respectively, at times of eruptive flares.
The spread of critical heights at CME onsets, and the observation of CMEs both when the critical height is rising and falling, demonstrates that knowledge of the critical height in an active region is not wholly sufficient to predict an eruption. 
Determining whether a magnetic flux rope has formed and/or risen to a height where the torus instability can occur is equally important. Therefore, more work should be done towards understanding these pre-eruptive magnetic structures. 
Furthermore, the torus instability may not drive the eruptions of all observed CMEs. The role of other mechanisms, such as magnetic breakout reconnection, should therefore still be considered when determining the mechanisms involved in the initiation of CMEs.

In combination with observational analysis of the magnetograms used, we see that the different types of phases correspond to different physical conditions in the solar atmosphere. When an increase in magnetic flux is related to the emergence of a new bipole into the corona (either into an area of quiet-Sun or a previously-established active region), we often see a cotemporal decrease in the local critical height. Conversely, when increasing magnetic flux is associated with the continuing emergence and expansion of a bipole that is already well-established in the atmosphere, we often observe a cotemporal increase in the critical height.
Phases of increasing unsigned magnetic flux and decreasing critical height were found to have relatively short durations, at 30.9 hours on average, whereas phases of increasing unsigned flux and increasing critical height lasted for an average of 42.2 hours. 
This difference in average phase duration highlights the contrasting physical scenarios described above, with the earliest phases of new bipoles emerging into equilibrium coronal magnetic fields causing relatively rapid decreases in critical height, whereas more gradual, continuing emergences of bipoles that already dominate the coronal field cause steadier rises in critical height as their associated coronal loops expand.

Further support to the distinct physical scenarios described above comes from examining the separation of magnetic polarities in each active region over time. \citet{torok2007simulations} show that the critical height increases in bipolar regions with larger separations between each pole, and in our study, we often see the critical height increasing (decreasing) at the same time as the separation of active region polarities increases (decreases).
When new magnetic flux emerges at the central PIL of an active region, for example between two existing sunspots, the mean separation of positive and negative flux in the area decreases. Furthermore, the newly emerged flux creates a sharp gradient of magnetic field strength low down in the solar atmosphere, meaning critical values of the decay index occur at lower heights. 
On the other hand, as flux continues to emerge, polarities separate from each other and magnetic loops expand higher into the corona. The vertical gradient in magnetic field strength lessens as the bipole's magnetic loops fill higher into the corona (where there had previously been relatively weak field), causing the critical height to increase.
When magnetic flux is decreasing, the mean separation between magnetic polarities sometimes increases (e.g. as sunspots disperse due to supergranular flows) and sometimes decreases (as sunspots converge and undergo flux cancellation), and the critical height also tends to change in the same way as the separation.

Since the critical height in an active region is related to the separation between magnetic polarities, and the emergence and decay of magnetic flux can cause the separations between polarities to change, the evolutions of magnetic flux and critical height are not independent of each other. This complexity should be acknowledged when attempting to apply the findings of our study to the estimation of CME likelihood. However, the result that, in our sample of 42 active regions, CME rates were 63\% higher during phases of increasing flux and increasing critical height than during phases of increasing flux and decreasing critical height is still remarkable.

Our findings support and extend the results of \citet{wang2017critical} and \citet{baumgartner2018eruptive}, who showed that, during two-ribbon solar flares, the active region critical height is proportional to the mean separation between magnetic polarities (with slopes of 0.54 and $0.4 \pm 0.1$, respectively). 
In this work, we take the mean of the critical height and the polarity separation over the observed life of each active region, and across the regions where we are most confident in our methodology (bipolar regions), we find a slope of $0.52 \pm 0.04$ between the quantities, in good agreement with the findings of the previous studies. This demonstrates how the proportionality between polarity separation and critical height holds well over time, and not only at the onsets of flaring events. Furthermore, the critical height in an active region may be approximated at any time as half of the separation between its magnetic polarities.

\section{Conclusions} \label{sec:conclusions}

In our sample of 42 active regions, observed CME rates were twice as high during phases when the unsigned magnetic flux through the photosphere was increasing rather than decreasing. Separately, CME rates were slightly (36\%) higher during phases when the critical height for the onset of the torus instability was increasing than when it was decreasing. Examining only the phases when magnetic flux was increasing, there were significantly (63\%) more CMEs when the critical height was rising than when it was falling. 
Changes in critical height are not independent of changes in magnetic flux, and yet our results suggest that knowledge of the critical height in combination with observations of emerging or decaying magnetic flux could help forecasters of space weather to determine the relative likelihood that certain active regions are more likely to produce CMEs during a given time interval than others.

Even without directly quantifying the magnetic decay index in the solar corona, the critical height in an active region can, on average be well-approximated as approximately half of the separation of its magnetic polarities throughout its lifetime. 
Furthermore, changes in this separation, and therefore changes in the critical height, can be well-understood in the context of the evolution of the total magnetic flux in an active region. The emergence of new, compact magnetic bipoles into quiet-Sun regions or at the central polarity inversion lines of existing active regions tends to cause the critical height to decrease, whereas the continuing emergence and expansion of established bipoles causes the critical height to increase.

We find that the critical height at CME onset most commonly lies between $40\ \mathrm{Mm}$ and $50\ \mathrm{Mm}$ above the photosphere, with a mean of $43 \pm 8\ \mathrm{Mm}$.
However, values ranging from $23 \pm 7\ \mathrm{Mm}$ to $67 \pm 15\ \mathrm{Mm}$ are noted, and we observe CMEs both at times when the critical height is falling and rising. This suggests that knowing only the critical height in an active region is insufficient for predicting CMEs, and therefore more work must be done to identify and understand the heights of pre-eruptive magnetic flux ropes in the solar atmosphere if we are to routinely predict their eruptions as CMEs.

\begin{acknowledgements}
A.W.J. is supported by a European Space Agency (ESA) Research Fellowship.
J.O. thanks the STFC for support via funding given in her PhD studentship.
We thank the anonymous referee for their constructive and important feedback on this work during the review process.
Data courtesy of NASA/SDO and the AIA and HMI science teams.
STEREO is the third mission in NASA’s Solar Terrestrial Probes program.
The SOHO LASCO CME catalog is generated and maintained at the CDAW Data Center by NASA and The Catholic University of America in cooperation with the Naval Research Laboratory. SOHO is a project of international cooperation between ESA and NASA.
This research has made use of SunPy v2.0.1 \citep{sunpy_2.0.1}, an open-source and free community-developed solar data analysis Python package \citep{sunpy2020}.
Many of the computed data outputs, including magnetic fluxes, critical heights, and polarity separations for each HARP region can be found at \url{https://github.com/alexjsolar/James_et_al_hc_evolution}, alongside example Python code for handling and plotting the data. 
\end{acknowledgements}

\bibliographystyle{aa}

\end{document}